\documentclass[twocolumn,seceq]{jpsj3}
\usepackage{graphicx}

\def\runauthor{T. {\sc Iharagi}, A. {\sc Gendiar}, 
H. {\sc Ueda}, 
and T. {\sc Nishino}}

\title{Phase Transition of the Ising model on a Hyperbolic Lattice}

\author
{Takatsugu {\sc Iharagi}$^{1)}$, Andrej {\sc Gendiar}$^{2,3)}$, 
Hiroshi {\sc Ueda}$^{4)}$, 
and Tomotoshi {\sc Nishino}$^{1)}$}
\inst
{$^1$Department of Physics, Graduate School of Science, 
Kobe University, Kobe 657-8501, Japan\\
$^2$Institute of Electrical Engineering, Slovak Academy of Sciences,
D\'ubravsk\'a cesta 9, SK-841 04, Bratislava, Slovakia\\
$^3$Institute of Physics, Slovak Academy of Sciences,
D\'ubravsk\'a cesta 9, SK-845 11, Bratislava, Slovakia\\
$^4$Department of Material Engineering Science, Graduate School of 
Engineering Science, Osaka University, Osaka 560-8531, Japan
}

\abst
{
The matrix product structure is considered on a regular lattice in the hyperbolic plane.
The phase transition of the Ising model is observed on the hyperbolic $( 5, 4 )$ lattice 
by means of the 
corner-transfer-matrix renormalization group (CTMRG) method. Calculated correlation 
length is always finite even at the transition temperature, where mean-field like behavior 
is observed. The entanglement entropy is also always finite. }

\kword{DMRG, CTMRG, Hyperbolic, Entanglement}

\begin{document}
\sloppy
\maketitle

\section{Introduction}

Classification of phase transitions is one of the central issue in the study of 
lattice models in statistical mechanics. When a system exhibits the second-order
transition, normally the correlation length diverges at the transition point.
As a result of scale invariance at the criticality, the transitions are characterized by scaling indices, 
where their values are completely classified in two-dimension by means of the conformal field 
theory. 

The mean-field like 2nd-order transition is exceptional in the point that the correlation length 
may not play an important role, in particular when the transition is described by the 
Landau free energy that is expressed as a simple polynomial of the order 
parameter.~\cite{Baxter1} In this article we focus on the mean-field like transition
observed for the Ising model on the hyperbolic lattices,~\cite{Rietman,Chris1,
Chris2,dAuriac,Doyon,Shima,Hasegawa,Ueda,Roman,Roman2}
the regular lattice in two-dimensional (2D) plane with constant negative curvature.~\cite{Sausset}
Among the hyperbolic $( p, q )$-lattices, which are the tessellations of the regular
$p$-gons with the coordination number $q$,~\cite{pq} we consider the  $( 5, 4 )$-lattice 
shown in Fig.~1 as an example. We calculate the correlation length $\xi$ and 
entanglement entropy $S$ in the neighborhood of the second-order transition 
temperature $T_0^{~}$, and judge whether or not the system is critical at this temperature.

In the next section we explain the matrix product structure of the Ising model on the 
$( 5, 4 )$-lattice. We employ the corner transfer matrix renormalization group (CTMRG) 
method,~\cite{Nishino2,Nishino3} 
a variant of the density matrix renormalization group (DMRG) 
method~\cite{White1,White2,Schollwoeck} 
applied to 2D classical models,~\cite{Nishino1} to obtain the thermodynamic
properties of the model. We show the calculated results on $\xi$ and $S$ in \S 3. 
Conclusions are summarized in the last section.

\begin{figure}
\centerline{\includegraphics[width=6.5cm,clip]{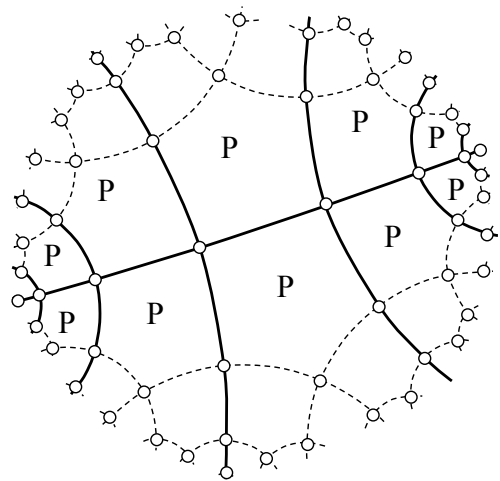}}
\caption{The hyperbolic $( 5, 4 )$ lattice drawn in the Poincar\'e disk. 
Open circles denote lattice points, where there are Ising spin variables.
Those regions denoted by $P$ correspond to half-column transfer matrices.}
\label{f1}
\end{figure}

\section{Matrix Product Structure on the Hyperbolic Lattice}

Consider the ferromagnetic Ising model on the $( 5, 4 )$-lattice shown
in Fig.~1. Each pair of neighboring sites is on a geodesic, which is drawn 
either by a line that passes through the center of the disk or by an arc. 
When there is no external magnetic field, the Hamiltonian is given by
\begin{equation}
H = -J \sum_{\langle i j \rangle}^{~} \sigma_i^{~} \sigma_j^{~} \, ,
\end{equation}
where $\sigma_i^{~} = \pm 1$ denotes the Ising spin variable at the $i$-th site, and
where $J > 0$ is the coupling strength between neighboring pair of sites 
denoted by $\langle i j \rangle$. It is convenient to introduce
the {\it interaction-round-a-face} (IRF) Boltzmann weight 
\begin{eqnarray}
&&W_{ijk\ell m}^{~} = \\
&&~~~
\exp\left[ - \beta \frac{J}{2} (
\sigma_i^{~} \sigma_j^{~} + 
\sigma_j^{~} \sigma_k^{~} +
\sigma_k^{~} \sigma_{\ell}^{~} +
\sigma_{\ell}^{~} \sigma_m^{~} +
\sigma_m^{~} \sigma_i^{~} ) \right]
\nonumber
\end{eqnarray}
for each pentagon, where $i, j, k, \ell$ and $m$ denote the sites around it, and
where $\beta$ represents the inverse temperature.
The partition function of the system is then expressed as
\begin{equation}
Z = \sum_{\rm config.}^{~} \prod_{\langle ijk\ell m \rangle}^{~} W_{ijk\ell m}^{~} \, ,
\end{equation}
where the product is taken over all the pentagons denoted by the 
group of sites $\langle ijk\ell m \rangle$,
and where the sum is taken over all the possible spin configurations.
We assume that the system is sufficiently large, and we consider the thermal equilibrium 
state deep inside the system, where the part is far from the system boundary.

Let us observe the structure of the $( 5, 4 )$-lattice. 
The thick straight ({\it horizontal}) line drawn in Fig.~1 divides the whole lattice into upper and
lower halves. Let us introduce the new labeling 
\begin{equation}
\ldots, \sigma_{\ell+1}^{(0)}, \sigma_{\ell}^{(0)}, 
\sigma_{\ell-1}^{(0)}, \sigma_{\ell-2}^{(0)}, \ldots
\end{equation}
from left to right for those spins on this line. We use the notation $\sigma^{(0)}_{\ldots}$ 
when we refer to the whole part of this {\it horizontal row} of spins. 

We next observe the thick arc that is perpendicular to the horizontal line we have considered,  
and that passes through the $\ell$-th site $\sigma_{\ell}^{(0)}$. We label those
spins on this {\it vertical} geodesic as
\begin{equation}
\ldots, \sigma_{\ell}^{(-1)}, \sigma_{\ell}^{(0)}, 
\sigma_{\ell}^{(1)}, \sigma_{\ell}^{(2)}, \ldots
\end{equation}
from downward to upward. We use the notation $\sigma^{\, \ldots}_{\ell}$ when we 
refer to the whole part of this {\it vertical column} of spins. For the latter convenience, we 
introduce the half-infinite spin columns
\begin{eqnarray}
\sigma^{\, \rm D}_{\ell} &=& 
\ldots, \sigma_{\ell}^{(-3)}, \sigma_{\ell}^{(-2)}, \sigma_{\ell}^{(-1)} \, \nonumber\\
\sigma^{\, \rm U}_{\ell} &=& 
\sigma_{\ell}^{(1)}, \sigma_{\ell}^{(2)}, \sigma_{\ell}^{(3)}, \ldots 
\end{eqnarray}
for the lower and the upper part of the column spin $\sigma^{\, \ldots}_{\ell}$.
It should be noted that the  spins
\begin{equation}
\ldots, \sigma_{\ell+1}^{(j)}, \sigma_{\ell}^{(j)}, 
\sigma_{\ell-1}^{(j)}, \sigma_{\ell-2}^{(j)}, \ldots
\end{equation}
for $j \neq 0$ are not on a geodesic, and therefore they 
cannot be regarded as a row spin. We have not explicitly shown
the system boundary, and have not given any special label 
to the boundary spins. When the system is off critical, we do
not have to consider about the system boundary so strictly.

We introduce the column-to-column transfer matrix
\begin{equation}
T_{\ell+1, \ell}^{~} 
=
T( \sigma^{\, \ldots}_{\ell+1} |  \, \sigma^{\, \ldots}_{\ell} ) \, ,
\end{equation}
which is created by multiplying all the IRF weights inside the {\it stripes} 
between $\ell+1$-th and $\ell$-th {\it vertical} geodesics, and taking
spin configuration sum except for those column spins  $\sigma^{\, \ldots}_{\ell+1}$ and 
 $\sigma^{\, \ldots}_{\ell}$. 
According to the division of the column spin
\begin{equation}
\sigma^{\, \ldots}_{\ell} = 
\sigma^{\, \rm D}_{\ell}, \, \sigma_{\ell}^{(0)}, \, \sigma^{\, \rm U}_{\ell} \, ,
\end{equation}
we can also express the column to column transfer matrix as a product of its 
upper and lower parts
\begin{eqnarray}
T_{\ell+1, \ell}^{~}  
&=& 
T( \sigma^{\, \rm D}_{\ell+1}, \sigma_{\ell+1}^{( 0 )}, \sigma^{\, \rm U}_{\ell+1} |  \, 
\sigma^{\, \rm D}_{\ell}, \sigma_{\ell}^{( 0 )}, \sigma^{\, \rm U}_{\ell} ) \nonumber\\
&=&
P_{\ell+1, \ell}^{\, \rm D} \, P_{\ell+1, \ell}^{\, \rm U} \, ,
\end{eqnarray}
where  $P_{\ell+1, \ell}^{\, \rm D}$ and $P_{\ell+1, \ell}^{\, \rm U}$ are half-column
transfer matrices (HCTMs)
\begin{eqnarray}
P_{\ell+1, \ell}^{\, \rm D} &=& 
P( \sigma^{\, \rm D}_{\ell+1}, \sigma_{\ell+1}^{(0)} |  \, 
\sigma^{\, \rm D}_{\ell}, \sigma_{\ell}^{(0)} )
\nonumber\\
P_{\ell+1, \ell}^{\, \rm U} &=&
P( \sigma_{\ell+1}^{(0)}, \sigma^{\, \rm U}_{\ell+1} |  \, 
\sigma_{\ell}^{(0)}, \sigma^{\, \rm U}_{\ell} ) \, .
\end{eqnarray}
Using these notations the lower and the upper halves of the system are 
represented by the IRF-type matrix products~\cite{IRF}
\begin{eqnarray}
\Psi^{\rm D}_{~}( \sigma_{\ldots}^{(0)} ) &=& 
\sum_{{\rm config.} \atop {\rm D}}^{~} \prod_\ell^{~} 
P( \sigma^{\, \rm D}_{\ell+1}, \sigma_{\ell+1}^{(0)} |  \, 
\sigma^{\, \rm D}_{\ell}, \sigma_{\ell}^{(0)} )
\nonumber\\
\Psi^{\rm U}_{~}( \sigma_{\ldots}^{(0)} ) &=& 
\sum_{{\rm config.} \atop {\rm U}}^{~} \prod_\ell^{~} 
P( \sigma_{\ell+1}^{(0)}, \sigma^{\, \rm U}_{\ell+1} |  \, 
\sigma_{\ell}^{(0)}, \sigma^{\, \rm U}_{\ell} ) \, ,\nonumber\\
\end{eqnarray}
respectively, where configuration sums are taken for all the $\sigma^{\, \rm D}_{\ell}$ 
and $\sigma^{\, \rm U}_{\ell}$. Note that the partition function $Z$ in Eq.~(2.3) is
expressed as the inner product between $\Psi^{\rm D}_{~}$ and $\Psi^{\rm U}_{~}$ 
\begin{equation}
Z = \sum_{\sigma_{\ldots}^{(0)} }^{~} 
\Psi^{\rm D}_{~}( \sigma_{\ldots}^{(0)} ) \, 
\Psi^{\rm U}_{~}( \sigma_{\ldots}^{(0)} ) \, ,
\end{equation}
where the configuration sum is taken over the horizontal row spins.

Statistical property of the left half  of the system can be represented by the 
successive product of the column-to-column transfer matrices
\begin{equation}
\Phi^{\rm L}_{~}( \sigma^{\, \rm D}_{0}, \sigma_{0}^{( 0 )}, \sigma^{\, \rm U}_{0} ) 
=
{\bf V}^{\rm L}_{~} \, \left( \prod_{\ell \ge 0}^{~} T_{\ell+1,\ell}^{~} \right) \, ,
\end{equation}
where ${\bf V}^{\rm L}_{~}$ is a vector that specifies the boundary condition at the
left border of the system. 
In the same manner the property of the right half of the system is represented as
\begin{equation}
\Phi^{\rm R}_{~}( \sigma^{\, \rm D}_{0}, \sigma_{0}^{( 0 )}, \sigma^{\, \rm U}_{0} ) 
=
\left( \prod_{\ell \le 0}^{~} T_{\ell,\ell-1}^{~} \right) {\bf V}^{\rm R}_{~} \, .
\end{equation}
We assume that a very weak symmetry breaking field is imposed to the
boundary spins, though we do not explicitly refer to this condition in the following.
Since the lattice structure shown in Fig.~1 is isotropic, it is also possible to 
represent $\Phi^{\rm L}_{~}( \sigma_0^{\, \ldots} )$ and 
 $\Phi^{\rm R}_{~}( \sigma_0^{\, \ldots} )$ as products of the 
 half-row transfer matrices in the
same manner as we have expressed $\Psi^{\rm D}_{~}( \sigma_{\ldots}^{(0)} )$ 
and $\Psi^{\rm U}_{~}( \sigma_{\ldots}^{(0)} )$ in Eq.~(2.12).~\cite{iTEBD}

Taking a partial contraction between $\Phi^{\rm L}_{~}( \sigma_0^{\, \ldots} )$ 
and $\Phi^{\rm R}_{~}( \sigma_0^{\, \ldots} )$, we
obtain the density matrix
\begin{eqnarray}
\rho^{\rm D}_{~} 
&=& \rho( {\bar \sigma}_{0}^{\rm D} |  \sigma_{0}^{\rm D} ) \\
&=& \sum_{ \sigma_{0}^{(0)} \, \sigma_{0}^{\rm U}}^{~}
\Phi^{\rm L}_{~}( {\bar \sigma}^{\, \rm D}_{0}, \sigma_{0}^{( 0 )}, \sigma^{\, \rm U}_{0} ) \, 
\Phi^{\rm R}_{~}( \sigma^{\, \rm D}_{0}, \sigma_{0}^{( 0 )}, \sigma^{\, \rm U}_{0} ) 
\nonumber
\end{eqnarray}
that is treated in the context of the bipartite entanglement to the 
vertical direction, or the block diagonal density matrix
\begin{eqnarray}
&&\rho( {\bar \sigma}_{0}^{\rm D} \sigma_{0}^{(0)} |  
\sigma_{0}^{\rm D} \sigma_{0}^{(0)} ) \\
&& ~~~~~~ = \sum_{ \sigma_{0}^{\rm U}}^{~}
\Phi^{\rm L}_{~}( {\bar \sigma}^{\, \rm D}_{0}, \sigma_{0}^{( 0 )}, \sigma^{\, \rm U}_{0} ) \, 
\Phi^{\rm R}_{~}( \sigma^{\, \rm D}_{0}, \sigma_{0}^{( 0 )}, \sigma^{\, \rm U}_{0} ) 
\nonumber
\end{eqnarray}
that is the 4-th power of the corner transfer matrix (CTM).~\cite{CTM1,CTM2,CTM3,Baxter1}

\section{Correlation Length Obtained by CTMRG}

The matrix product representation of $\Psi^{\rm D}_{~}( \sigma_{\ldots}^{(0)} )$ 
and $\Psi^{\rm U}_{~}( \sigma_{\ldots}^{(0)} )$ in Eq.~(2.12) has the same form as 
those for the square lattice Ising model. Thus we can apply Baxter's variational 
formulation~\cite{CTM1,CTM2,CTM3,Baxter1}
or the DMRG method~\cite{White1,White2,Nishino1,Schollwoeck} 
for the calculation of the free energy and other thermodynamic functions.
In this article we employ the CTMRG method,~\cite{Nishino2,Nishino3} 
a variant of the DMRG method, to obtain the correlation length $\xi$ and the 
entangle entropy $S$ of the Ising model on the $( 5, 4 )$ lattice. 

The CTMRG method maps the half-column spins $\sigma_\ell^{\rm D}$ and 
 $\sigma_\ell^{\rm U}$, respectively, to block spins $\zeta_\ell^{\rm D}$ and
 $\zeta_\ell^{\rm U}$ by means of the the renormalization group (RG) transformation, 
which is obtained by the diagonalization of the density matrix in Eq.~(2.17), or 
that in Eq.~(2.16).
Through this RG transformation, the HCTMs are mapped to the renormalized ones
\begin{eqnarray}
{\tilde P}^{\rm D}_{\ell+1,\ell} &=&
{\tilde P}( \zeta^{\rm D}_{\ell+1} , \sigma^{(0)}_{\ell+1} |  \, 
\zeta^{\rm D}_{\ell} , \sigma^{(0)}_{\ell} ) \nonumber\\
{\tilde P}^{\rm U}_{\ell+1,\ell} &=&
{\tilde P}( \sigma^{(0)}_{\ell+1}, \zeta^{\rm U}_{\ell+1} |  \, 
\sigma^{(0)}_{\ell} , \zeta^{\rm U}_{\ell} ) \, ,
\end{eqnarray}
where we put ``$\tilde~$'' marks on top of renormalized matrices. 
The column-to-column transfer matrix is renormalized in the same manner
\begin{eqnarray}
{\tilde T}_{\ell+1,\ell}^{~} &=& 
{\tilde T}( \zeta^{\, \rm D}_{\ell+1}, \sigma_{\ell+1}^{( 0 )}, \zeta^{\, \rm U}_{\ell+1} |  \, 
\zeta^{\, \rm D}_{\ell}, \sigma_{\ell}^{( 0 )}, \zeta^{\, \rm U}_{\ell} ) \nonumber\\
&=&
{\tilde P}^{\rm D}_{\ell+1,\ell} \, {\tilde P}^{\rm U}_{\ell+1,\ell} \, .
\end{eqnarray}
In the previous studies we have shown that the system exhibits the 
mean-field like second-order phase transition,~\cite{Ueda,Roman,Roman2}
where the spontaneous magnetization 
\begin{equation}
M = 
\frac{ \displaystyle \sum_{\zeta_{0}^{\rm D} \sigma_{0}^{(0)} }^{~}
\sigma_{0}^{(0)} 
\rho( 
\zeta_{0}^{\rm D} \sigma_{0}^{(0)} |  \, 
\zeta_{0}^{\rm D} \sigma_{0}^{(0)} )
}{ \displaystyle \sum_{\zeta_{0}^{\rm D} \sigma_{0}^{(0)} }^{~}
\rho( 
\zeta_{0}^{\rm D} \sigma_{0}^{(0)} |  \, 
\zeta_{0}^{\rm D} \sigma_{0}^{(0)} )
}
\end{equation}
 below the transition temperature $T_0^{~}$ 
is proportional to $\sqrt{T_0^{~} - T}$.

\begin{figure}
\centerline{\includegraphics[width=6.5cm,clip]{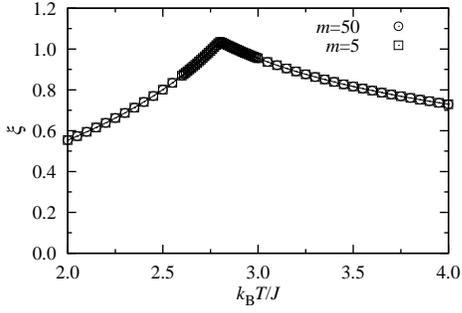}}
\caption{ Correlation Length $\xi$ with respect to the temperature $T$.}
\label{f2}
\end{figure}

Figure 2 shows the correlation length
\begin{equation}
\xi = \frac{1}{\log \lambda_0^{~} - \log \lambda_1^{~} } 
\end{equation}
calculated from the largest eigenvalue $\lambda_0^{~}$ and the second
largest one $\lambda_1^{~}$ of ${\tilde T}_{\ell+1,\ell}^{~}$. 
We have regarded the interaction parameter $J$ as the unit of energy, and
set the lattice constant as the unit of length. We keep 
at most $m = 40$ states for block spins, and actually $m = 5$ is sufficient
enough to draw the figure. It is clear that $\xi$ is of the order of the lattice constant 
even at the second order transition point $T_0^{~} = 2.799$. The fact  shows 
that the system is always off-critical, and thus the transition point cannot be
called as the {\it critical} point. The entanglement entropy
\begin{equation}
S = - \sum_i^{~} \omega_i^{~} \log \omega_i^{~} \, ,
\end{equation}
where $\omega_i^{~}$ is the $i$-th eigenvalue of the reduced density
matrix ${\tilde \rho}^{\rm D}_{~} = {\tilde \rho}^{\rm D}_{~}( {\bar \zeta}^{\rm D}_0 |  \, 
\zeta^{\rm D}_0 )$ in Eq.~(2.16), shown in Fig.~3 is also finite for any temperature $T$. 
The fact coincides that a very small number of states $m = 5$ is sufficient for
getting thermodynamic quantities precisely.~\cite{Ueda,Roman,Roman2}

\begin{figure}
\centerline{\includegraphics[width=6.5cm,clip]{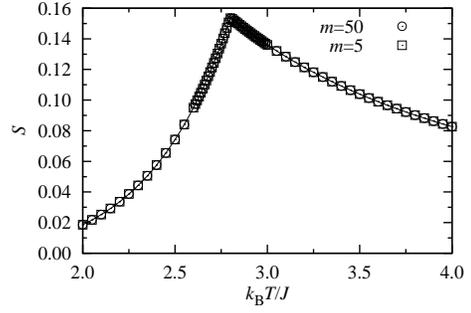}}
\caption{ Entanglement entropy of the MPS in Eq.~(2.12).  }
\label{f3}
\end{figure}

\section{Conclusions and Discussions}

We have calculated the correlation function $\xi$ and the entanglement entropy $S$ 
of the Ising model on the hyperbolic $( 5, 4 )$ lattice. Both of them remains finite at 
the transition temperature $T_0^{~}$. Therefore the mean-field like second-order
phase transition observed for this system is not related to critical phenomena with
diverging $\xi$. 
This off-critical behavior is common to phase transitions observed in the
tensor product formulations,~\cite{Kramers,3D,3D2,3D3,Liu} where the trial state is
finitely correlated due to the state number limitation for the block spin variables.

The mean-field nature of the phase transition might be explained 
by the path integral representation of the Green function on the lattice, 
if dominant contribution comes from the shortest path between two points. 
It should be noted that in the hyperbolic plane any deviation from the 
shortest path causes increase of path length more than that on the flat plane.
Such an increase reduces entanglement between two points. It should
be noted that the Hausdorff dimension of the $( 5, 4 )$ lattice, or more general 
the hyperbolic $( p, q )$ lattices, is infinite. 

From the path integral picture on the hyperbolic plane, one reaches a deformation to 
1D quantum Hamiltonians, so called the hyperbolic deformation.~\cite{HUeda,HUeda2}
We conjecture that mean-field behavior also appears in ground-state phase
transitions of deformed 1D quantum systems,
when $N \lambda$ is sufficiently larger than unity, where $N$ is the system size.

It is known that Kasteleyn formalism is applicable for dimer models on the hyperbolic 
lattices.~\cite{Lund} Thus there is a mathematical interest on the hyperbolic lattices 
to find out a commutable row-to-row or column-to-column transfer matrices.

\section*{Acknowledgement}

The authors thank to A.~Sandvik for valuable discussions on the phase transition
in tensor product formulations.
A.~G. acknowledges the support of ERDF OP R\&D, Project
hQUTE - Centre of Excellence for Quantum Technologiesh
(ITMS 26240120009), CE QUTE SAV, APVV-51-003505, VVCE-0058-07, and VEGA-2/0633/09.
T.~N. thanks to Y.~Hayashi for informing us the 
Kasteleyn formalism on the hyperbolic lattices.
This work was partially supported by Grant-in-Aid for Scientific Research (C).


\begin{thebibliography}{99}
\bibitem{pq} When $( p - 2 ) (q - 2 )$ is larger than 4, the lattice is on the hyperbolic plane.
\bibitem{IRF} It is possible to perform the DMRG calculation, assuming that the
variational state is expressed as the IRF-type matrix product, as proposed by Baxter.
\bibitem{iTEBD} The vertical and horizontal direction is equivalent for the isotropic Ising
model on the square lattice. Thus the iTEBD method, which assumes horizontal MPS, 
applied to the square lattice Ising model works almost equivalent as the infinite system 
DMRG method, which creates vertical MPS.
\end{thebibliography}
\end{document}